\documentclass[10pt]{article}
\usepackage{graphicx}
\usepackage{ulem}

\def\Title#1{\begin{center} {\Large #1 } \end{center}}
\def\Author#1{\begin{center}{ \sc #1} \end{center}}
\def\Address#1{\begin{center}{ \it #1} \end{center}}

\newcommand\pubblock{\rightline{\begin{tabular}{l} Proceedings of the Second Annual LHCP\\ \pubnumber\\
         \pubdate  \end{tabular}}}

\newenvironment{Abstract}{\begin{quotation} \begin{center} 
             \large ABSTRACT \end{center}\bigskip 
      \begin{center}\begin{large}}{\end{large}\end{center} \end{quotation}}

\newenvironment{Presented}{\begin{quotation} \begin{center} 
             PRESENTED AT\end{center}\bigskip 
      \begin{center}\begin{large}}{\end{large}\end{center} \end{quotation}}

\def\Acknowledgements{\bigskip  \bigskip \begin{center} \begin{large}
             \bf ACKNOWLEDGEMENTS \end{large}\end{center}}
\def\beq{\begin{equation}}
\def\eeq#1{\label{#1}\end{equation}}
\def\eeqn{\end{equation}}

\def\beqa{\begin{eqnarray}}
\def\eeqa#1{\label{#1}\end{eqnarray}}
\def\eeqan{\end{eqnarray}}

\let\bar=\overbar

\def\Dslash{\not{\hbox{\kern-4pt $D$}}}
\def\dslash{\not{\hbox{\kern-2pt $\del$}}}

\def\msb{{\bar{\ssstyle M \kern -1pt S}}}

\usepackage{color}

\newcommand\pubnumber{KCL-PH-TH/2014-35, LCTS/2014-34 \\ CERN-PH-TH/2014-161}
\newcommand\pubdate{\today}

\def\affiliation{
Theoretical Particle Physics and Cosmology Group, Department of Physics, \\
King's College London, Strand, London WC2R 2LS, U.K; \\
Theory Division, Physics Department, CERN, CH 1211 Geneva 23, Switzerland}

\begin{document}

% large size for the first page
\large
\begin{titlepage}
\pubblock

%% Change the title, name, abstract
%% Title 
\vfill
\Title{  Theory Summary and Prospects  }
\vfill

%  if you need to add the support use this, fill the \support definition above. 
%   \Author{ FIRSTNAME LASTNAME \support }
\Author{ John ELLIS }
\Address{\affiliation}
\vfill
\begin{Abstract}

This talk reviews some of the theoretical progress and outstanding issues in QCD, \\ flavour
physics, Higgs and electroweak physics and the search for physics beyond
the Standard Model at the Tevatron and the LHC, and previews some
physics \\ possibilities for future runs of the LHC and proposed future
hadron colliders.

\end{Abstract}
\vfill

% DO NOT CHANGE 
\begin{Presented}
The Second Annual Conference\\
 on Large Hadron Collider Physics \\
Columbia University, New York, U.S.A \\ 
June 2-7, 2014
\end{Presented}
\vfill
\end{titlepage}
\def\thefootnote{\fnsymbol{footnote}}
\setcounter{footnote}{0}
%

% normal size for the rest
\normalsize 

%% Your paper should be entered below. 

\section{Introduction}

The year 2014 is a year of anniversaries. In the world of physics,
it was 150 years ago that Maxwell announced the unification of electricity and magnetism. 
On the other hand, World War 1 started 100 years ago, 
World War 2 started in Europe 75 years ago, and D-Day was 70 years ago.
Back in the world of physics, Feyman diagrams were formulated 65 years ago,
and $\pi^0 \to \gamma \gamma$ decay was calculated by Jack Steinberger. In 1954,
60 years ago, Peter Higgs obtained his PhD, CERN was founded, and Fermi extrapolated
future accelerators to the circumference of the Earth. 50 years ago, 1964 was a bumper
year: quarks were postulated, as were charm and colour, spontaneous symmetry breaking
in gauge theories was discovered, Bell's theorem was proved, the $\Omega^-$, CP violation
and the cosmic microwave background were discovered. Meanwhile, in the larger world the
 civil rights legislation was passed in the US, and the Beatles invaded. Supersymmetric
 field theories in four dimensions were first proposed 40 years ago, and the year 1974
 also witnessed the November revolution when the $J/\psi$ particle was discovered.
 In 1979, 35 years ago, the gluon was discovered. The first LHC workshop was held 30 years ago
 in 1984, and supersymmetry was not discovered for the first time at the CERN $p - \bar{p}$ collider.
25 years ago, 1989 saw the invention of the World-Wide Web, as well as the downfall of (much of) communism.
The approval of the LHC came 20 years ago, in 1994, and the first low-energy collisions took place
5 years ago, in 2009.

This conference has provided an opportunity to review the current landmarks in collider physics~\cite{JV}.
What should we remember from 2014?~\footnote{The references are largely to plenary talks at
the conference: I apologize to those I omit, to the uncited parallel session speakers, and to the authors of
many uncited original papers.}

\section{QCD}

The foundation for all physics at the LHC is QCD~\cite{RB,MP}: it is the dominant force in particle production,
providing us with backgrounds and pile-up events~\cite{PA} as well as tests of the Standard Model, and
better understanding may enable us to dig out new possible signals, e.g., in boosted jets~\cite{EK}. Here
I will just discuss a few QCD topics: top physics, especially the top quark mass, the production of
new particles, especially the Higgs boson, $W^+ W^-$ production, and heavy-ion collisions.

The top quark mass $m_t$ is a basic parameter of the Standard Model, whose exact value is
crucial for the stability of the electroweak vacuum~\cite{AJ}. Understanding QCD accurately is crucial for
extracting from data a precise value of $m_t$ using some suitable theoretical definition related
to the underlying Lagrangian. However, simulations for comparison with experiment are typically calculated in terms of a
Monte Carlo mass whose relation to the pole or running mass is unclear.
%The running mass is in principle the best defined theoretically, but there is an estimated
%uncertainty of $\pm 0.7$~GeV in the relation between this and the Monte Carlo mass, and
%it is also estimated that there is a non-perturbative correction between the running mass
%and the pole mass that may be $0.5$~GeV. 

An experimental world average value of $m_t$ was
recently announced:
\begin{eqnarray}
{\rm World:} & m_t & = \; 173.34 \pm 0.36 \pm 0.67 \;  = \; 173.34 \pm 0.76~{\rm GeV} \, .
\label{mt}
\end{eqnarray}
Additionally,  two new measurements using lepton + jets final states were reported at this meeting:
\begin{eqnarray}
{\rm CMS~\cite{JV}:} & m_t & = \; 172.04 \pm 0.77~(0.19 \pm 0.75)~{\rm GeV} \, , \nonumber \\
{\rm D0~\cite{AJ}:} & m_t & = \; 174.98 \pm 0.76~(0.58 \pm 0.49)~{\rm GeV} \, ,
\label{newmt}
\end{eqnarray}
and CMS has subsequently reported a measurement $m_t = 172.08 \pm 0.90~(0.36 \pm 0.83)$ GeV 
using all-jet final states~\cite{CMStjets}.
It is essential that the theoretical QCD uncertainties in the Monte Carlo $\to$ running mass $\to$
pole mass corrections be reduced below the improving experimental precision. At present, it is
estimated that these corrections are~\cite{SM}
\begin{eqnarray}
{\rm Monte~Carlo} \to {\rm Running~mass} & \pm & 0.7~{\rm GeV} \, , \nonumber \\
{\rm Running~mass} \to {\rm Pole~mass} & + & 0.5~{\rm GeV} \, ,
\label{masscorrections}
\end{eqnarray}
which are not negligible. The first of these sources of uncertainty will be removed if the
measurement of the total $t \bar{t}$ production cross section can be used in the future~\cite{MSS},
but this will require significant reductions in the theoretical and experimental uncertainties.

Another instance where accurate higher-order QCD calculations are at a premium
is in the dominant gluon-fusion contribution to the Higgs production cross section~\cite{SD}. Several
different NNLO calculations are available, and are included in various publicly-available
tools. Unfortunately the agreement between them is unsatisfactory.
Fortunately, progress is being made on an NNNLO calculation.
This will improve the theoretical accuracy, but progress in convergence between the
PDFs will also be needed in order to reduce the theoretical uncertainties below the
experimental measurement uncertainties~\cite{MU}.

Overall, perturbative QCD calculations are doing a fantastic job of predicting the production cross sections
of jets, massive vector bosons~\cite{BJ,TK} and the Higgs boson at the LHC, the most striking exception being $W^+ W^-$
production~\cite{JH}:
\begin{eqnarray}
{\rm CMS~8~TeV:} & 69.9 \pm 2.8 \pm 5.6 \pm 3.1~{\rm pb} \; \; \; \; {\rm cf,~Theory:} & 57.3^{+2.3}_{-1.6}~{\rm pb}\, , \nonumber \\
{\rm ATLAS~7~TeV:} & 51.9 \pm 2.0 \pm 3.9 \pm 2.0~{\rm pb} \; \; \; \; {\rm cf,~Theory:} & 44.7^{+2.1}_{-1.9}~{\rm pb}
\label{WW}
\end{eqnarray}
This discrepancy has triggered some enthusiastic ambulance-chasing with scenarios involving a light
stop squark, chargino and bino~\cite{ambulance}. However, there is an ongoing debate whether the experimental
discrepancy with theory should yet be taken seriously: anomalous electroweak interactions seem
unable to explain the enhancements in the cross sections, but higher-order QCD corrections
predictions increase the $W^+ W^-$ cross section by several \%~\cite{NNLOWW},
and better understanding of jet resummation and the
modelling of the experimental jet vetos may also reduce the discrepancy~\cite{antiamb}.

\begin{figure}[htb]
\centering
\includegraphics[height=2.3in]{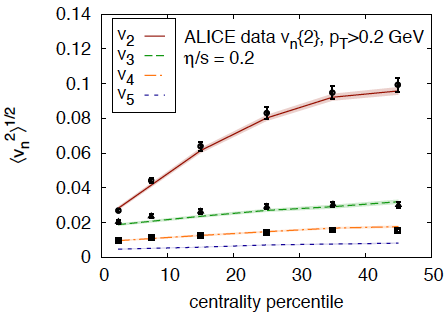}
\caption{ Measurements of the transverse flow parameters $v_n$ in lead-lead collisions at the LHC
are in good agreement with hydrodynamic simulations with a shear viscosity-to-entropy ratio $\eta/s \sim 0.2$~\protect\cite{Gale,WL}.}
\label{fig:eta}
\end{figure}

A very different QCD arena is provided by heavy-ion collisions, which probe the nature of
hot and dense quark-gluon matter~\cite{DT2}. Lattice calculations provides quantitative
understanding of the equation of state, and attention now focuses on its dynamical
properties. The transverse flow pattern in near-central collisions `remember' the
transverse shapes of the nuclei initiating each `Little Bang', 
and experimental measurements indicate~\cite{Gale,WL} that the quark-gluon medium behaves like a near-perfect
fluid, as illustrated in Fig.~\ref{fig:eta}. It has a very low shear-viscosity-to-entropy ratio $\eta/s$ lying within a factor $\sim 2$
of the lower bound $\eta/s = 1/4 \pi$ that holds in a wide class of strongly-interacting field theories,
as was first found using the AdS/CFT correspondence~\cite{Son} developed on the basis of holography
as formulated in the context of string theory~\footnote{An intriguing recent development is the discovery
of collective phenomena such as a nearside ridge in proton-lead collisions~\cite{AK}. The good news is
that such phenomena may enable the transverse structure of the proton to be probed: is it Y- or $\Delta$-shaped?~\cite{CSM}}.
Jet quenching provides another probe of quark-gluon matter~\cite{BC}: it seems that the `missing' energy
splashes out at relatively large angles to the jet axis, though interpretations in terms of Mach cones
or {\v C}erenkov radiations have not been confirmed. The production of quarkonia also probes the
quark-gluon medium: $J/\psi$ production is strongly suppressed at lower energies, as is
$\upsilon$ production at the LHC, presumably because of
Debye screening or dissociation in the medium. On the other hand, $J/\psi$ production recovers at LHC energies,
presumably because of $c \bar{c}$ regeneration in the medium~\cite{ES}.

%%%%%%%%%%%%%%%%%%%%%%%%%%%%%%%%%%%%%%%%%%%%%%%%%%%%%%%%%%%%%%%%%%%%%%%%%
%%
%%   use this format to include an .eps figure into your paper
%%
%%%%%%%%%%%%%%%%%%%%%%%%%%%%%%%%%%%%%%%%%%%%%%%%%%%%%%%%%%%%%%%%%%%%%%%%%%%

Relativistic heavy-ion collisions at RHIC and the LHC have certainly produced a medium with interesting
collective properties: its characterization is still a work in progress.

\section{Flavour Physics}

The Cabibbo-Kobayashi-Maskawa (CKM) description of flavour mixing and CP violation
in the quark sector also works very well~\cite{ZL}. It has made many successful predictions, including
many modes of CP violation in the $K^0$, $B^0$, $B^\pm$, and $B_s$ systems~\cite{YA}. A while back,
there was an indication of CP violation in $D^0$ decays at a level that would have been very
difficult to explain within the CKM picture, but this indication has not been confirmed
by additional data~\cite{YA}. Another success for the CKM model has been the prediction of the branching
ratio for the rare decay $B_s \to \mu^+ \mu^-$, which is confirmed by experiment at the 30\%
level, providing an interesting constraint on extensions of the Standard Model such as superymmetry.
A key prediction of CKM and related minimal flavour violation (MFV)
models that remains to be tested is the ratio of $B_d \to \mu^+ \mu^-/B_s \to \mu^+ \mu^-$.

\begin{figure}[htb]
\centering
\includegraphics[height=2.3in]{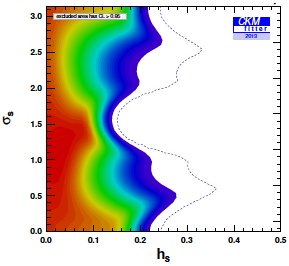}
\caption{ Experimental constraint on a possible non-Standard Model contribution to $B_s$ mixing~\protect\cite{CKMFitter}.}
\label{fig:hs}
\end{figure}

Despite these successes, there is still considerable scope for new physics beyond the CKM
paradigm. For example, there could still be a substantial BSM contribution to the mixing amplitude
for $B_s$ mesons: $A = A|_{SM} \times (1 + h_s e^{i \sigma_s})$,
as seen in Fig.~\ref{fig:hs}~\cite{CKMFitter}. It is still an open question whether any new TeV-scale physics must 
necessarily copy CKM slavishly {\it \`a la} MFV. Indeed, there a few anomalies to whet the appetite.
The CP asymmetries in $B^0 \to K^\pm \pi^\mp$ and $B^\pm \to K^\pm \pi^0$ are quite
different, though this most likely due to poorly-understood strong-interaction effects.
The branching ratio for $B^\pm \to \tau^\pm \nu$ differs from the Standard Model prediction
by $\sim 2 \sigma$, though this seems difficult to explain within 2-Higgs-doublet models or supersymmetry.
There is an $\sim 3.7 \sigma$ anomaly in the $P_5^\prime$ angular distribution for
$B^0 \to K^{*0} \mu^+ \mu^-$ that could be explicable by the contribution
of a $Z^\prime$ boson (though its significance is reduced to $\sim0.5 \%$ when the look-elsewhere
effect is taken into account). There are discrepancies in the determinations of the CKM matrix element
$V_{ub}$ that might be a signature of a vector-like quark. There is a persistent anomaly in the
diimuon asymmetry at the Tevatron~\cite{PG}. On the other hand, measurements of the $t \bar{t}$ 
forward-backward asymmetry now agree quite well with higher-order QCD calculations~\cite{PG}, as does the
corresponding $t \bar{t}$ rapidity asymmetry at the LHC.

%%%%%%%%%%%%%%%%%%%%%%%%%%%%%%%%%%%%%%%%%%%%%%%%%%%%%%%%%%%%%%%%%%%%%%%%%
%%
%%   use this format to include an .eps figure into your paper
%%
%%%%%%%%%%%%%%%%%%%%%%%%%%%%%%%%%%%%%%%%%%%%%%%%%%%%%%%%%%%%%%%%%%%%%%%%%%%

There are plenty of flavour issues to be addressed in future LHC runs, and at Super-KEKB.

\section{Higgs Measurements}

The mass of the Higgs boson is most accurately measured in the $\gamma \gamma$
and $Z Z^* \to 2 \ell^+ 2 \ell^-$ final states~\cite{BL,XJ}. For some time, there has been tension between
the masses measured in these channels by ATLAS. The new values reported at this conference are~\cite{DF}:
\begin{eqnarray}
H \to \gamma \gamma: m_H & = \; 125.98 \pm 0.42 \pm 0.28~{\rm GeV} = \; 125.98 \pm 0.50~{\rm GeV}\, , \nonumber \\
H \to Z Z^*: m_H & = \; 125.51 \pm 0.52 \pm 0.04~{\rm GeV} = \; 125.51 \pm 0.52~{\rm GeV}\, , \nonumber \\
{\rm ATLAS~combined:} ~m_H & = \; 125.36 \pm 0.37 \pm 0.18~{\rm GeV} \; = \; 125.36 \pm 0.41~{\rm GeV}\, ,
\label{ATLASm}
\end{eqnarray}
with a mass difference:
\begin{equation}
\Delta m_H \; = \; 1.47 \pm 0.67 \pm 0.18~{\rm GeV} = \; 1.47 \pm 0.72~{\rm GeV}\, ,
\label{ATLASDeltam}
\end{equation}
that has $\sim 2$-$\sigma$ significance. At the conference, CMS reported a precise mass only in the
$Z Z^*$ channel~\cite{MK}:
\begin{equation}
H \to Z Z^*: m_H \; = \; 125.6 \pm 0.4 \pm 0.2~{\rm GeV}\, ,
\label{CMSZZ*}
\end{equation}
but CMS has subsequently also reported a precise value in the $\gamma \gamma$ channel~\cite{CMSgaga}:
\begin{equation}
H \to \gamma \gamma: m_H \; = \; 124.70 \pm 0.31 \pm 0.15~{\rm GeV} \; = \; 124.70^{+0.35}_{-0.34}~{\rm GeV}\, .
\label{CMSgaga}
\end{equation}
Combining the two measurements (\ref{CMSZZ*}, \ref{CMSgaga}), one finds
\begin{equation}
{\rm CMS~combined:} ~m_H \; = \; 125.03^{+ 0.26}_{-0.27}~^{+ 0.13}_{- 0.15}~{\rm GeV} \; = \; 125.03 \pm 0.30~{\rm GeV}\, .
\label{CMSm}
\end{equation}
Amusingly, there is also some tension between the two CMS measurements (\ref{CMSZZ*}, \ref{CMSgaga}):
\begin{equation}
\Delta m_H \; = \; - 0.9 \pm 0.4 \pm 0.2 ^{+0.34}_{-0.35}~{\rm GeV} \, ,
\label{CMSDeltam}
\end{equation}
but it has the opposite sign from the ATLAS mass difference (\ref{ATLASDeltam})!
Combining naively the ATLAS and CMS measurements yields
\begin{equation}
m_H \; = \; 125.15 \pm 0.24~{\rm GeV}.
\label{mH}
\end{equation}
The precise value of $m_h$ is also important for the stability of the electroweak vacuum in the Standard Model,
as will be discussed later.

Before that, however, it is worth remembering that the fitted values of $m_h$ measured in
$\gamma \gamma$ and $Z Z^*$ final states {\it should} differ as a result of interference in the $\gamma \gamma$
final state between $H$ production and QCD production of $\gamma$ pairs. This effect depends
on the total width $\Gamma_H$ of the Higgs boson, but its magnitude is very small in the Standard
Model: $\Delta m_H = - 54$~MeV~\cite{SD}, which is far below the current precision and probably beyond the
reach of the high-luminosity LHC.

\begin{figure}[htb]
\centering
\includegraphics[height=2.3in]{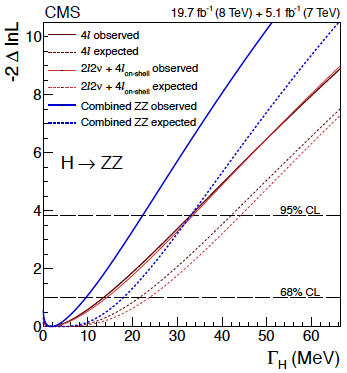}
\caption{ CMS constraint on $\Gamma_H$ from the off-shell Higgs contribution
to $ZZ$ production~\protect\cite{CMSGammaH}.}
\label{fig:CMSGammaH}
\end{figure}

The strongest bound on $\Gamma_H$ currently comes from constraints on the off-shell Higgs contribution
to $ZZ$ production~\cite{GammaHTH}. The pioneering measurement reported at this conference by CMS gave~\cite{CMSGammaH}
\begin{equation}
\Gamma_H \; < \; 5.4 \times \Gamma_H |_{SM} \, ,
\label{CMSGammaH}
\end{equation}
as seen in Fig.~\ref{fig:CMSGammaH},
which is to be compared with the Standard Model prediction $\Gamma_H = 4.2$~MeV. Subsequently
ATLAS has also reported an upper limit~\cite{ATLASGammaH}
\begin{equation}
\Gamma_H \; < \; 10.7 \times \Gamma_H |_{SM} \, ,
\label{ATLASGammaH}
\end{equation}
However, there is some model-dependence in the interpretation of these limits, since
some extensions of the Standard Model could also affect the $ZZ$ production rate~\cite{CEMS}.

%%%%%%%%%%%%%%%%%%%%%%%%%%%%%%%%%%%%%%%%%%%%%%%%%%%%%%%%%%%%%%%%%%%%%%%%%
%%
%%   use this format to include an .eps figure into your paper
%%
%%%%%%%%%%%%%%%%%%%%%%%%%%%%%%%%%%%%%%%%%%%%%%%%%%%%%%%%%%%%%%%%%%%%%%%%%%%

Combining their measurements
of the coupling strengths of the Higgs boson to $\gamma \gamma$,
$Z Z^*$, $W W^*$~\cite{BL,XJ}, $b \bar{b}$ and $\tau^+ \tau^-$~\cite{CG,JK}, ATLAS
and CMS report the following mean signal strengths:
\begin{eqnarray}
{\rm ATLAS:} ~\mu & = \; 1.30 \pm 0.12 \pm 0.10 \pm 0.09 & \, , \nonumber \\
{\rm CMS:} ~ \mu & = \; 1.00 \pm 0.09~^{+ 0.08}_{- 0.07} \pm 0.07 & = \; 1.00 \pm 0.13 \, .
\label{mu}
\end{eqnarray}
Overall, the available measurements are quite compatible with the Standard Model,
and there are no signs of any other Higgs bosons~\cite{WD,MM}. Measurements at
the Tevatron are also compatible with a Standard Model Higgs boson~\cite{WF}.

\section{Quo Vadis?}

Run 1 of the LHC has sown a great amount of theoretical confusion, since many
fashionable scenarios for physics beyond the Standard Model had predicted either that new
particles or other phenomena would be observed, or that the Higgs boson would either be
significantly different from that predicted in the Standard Model or that it would not exist at all.
Instead, there is no evidence for extra dimensions opening up, no signs that the Higgs boson may be
composite, and no indication of supersymmetry. There has been a mass extinction of theories, and the
survivors have had to evolve. Many theorists are asking whether our ideas of
naturalness should be modified or even abandoned. Does Nature care, or should we be happy if we
can survive somewhere in the string landscape? In the case of supersymmetry, for example, theorists have been
considering models with split or high-scale SUSY? My own point of view is that SUSY anywhere
is better than nowhere. On the other hand, it is clear that SUSY could not explain by itself the hierarchy of
mass scales in physics, and new ideas beyond SUSY are needed. To paraphrase that well-known
contemporary philosopher Lionel Messi {\it ``In \sout{football as in watchmaking} particle theory talent and elegance
mean nothing without rigour and precision"}~\cite{SD}.

Faced with the apparent completion of the Standard Model by the discovery of a (the?)
Higgs boson, one might be tempted to think there is no physics beyond the Standard Model,
but beware historical hubris! In 1894, Albert Michelson declared that {\it ``The more important
fundamental laws and facts of physical science have all been discovered"}, just three
years before the discovery of the electron. Not to be outdone, in 1900 Lord Kelvin declared that
{\it ``There is nothing new to be discovered in physics now,
all that remains is more and more precise measurement"}, just five years before Einstein
postulated the photon as an explanation of the photoelectric effect, explained Brownian
in terms of atoms and proposed special relativity.
More recently, in 1981 Stephen Hawking asked {\it ``Is the end of theoretical physics in sight?"}.

To paraphrase James Bond, there are plenty of reasons to think that the Standard Model is not enough~\cite{JB},
but let me just mention 007 of them. 1) ``Empty" space is {\it probably} unstable, if one takes at face value
the measured values of $m_t$ and $m_H$ discussed above, and there is no additional physics.
2) The Standard Model does not have a candidate for the dark matter required by astrophysics and cosmology.
3) The Standard Model does not explain the origin of matter in the Universe.
4) The Standard Model does not explain the small neutrino masses.
5) The Standard Model does not explain why the weak interactions are so strong relative to gravity (the hierarchy problem).
6) The Standard Model is (probably) not capable of inflating the Universe~\cite{AK2}, in particular 
because the effective Higgs potential is probably negative at high scales, as discussed in the next Section.
7) How does one make a consistent quantum theory of gravity?

In the following sections I discuss some of these.

\section{The Instability of the Electroweak Vacuum}

We think that the effective potential in the Standard Model resembles a Mexican hat,
rotationally symmetric, unstable at the origin, with by a circular valley 
where $\langle H \rangle \equiv v = 246$~GeV, and beyond it
a rising brim. But with the measured values of $m_t$ and $m_H$, it seems that
renormalization by the top quark turns the electroweak brim 
down at large field values, like an Australian bush-hat whose brim is
weighed down by dangling corks. This means that the desired electroweak vacuum with
$v = 246$~GeV is unstable with respect to quantum tunnelling though the brim, into an
anti-de-Sitter 'Big Crunch'.

\begin{figure}[htb]
\centering
\includegraphics[height=2.3in]{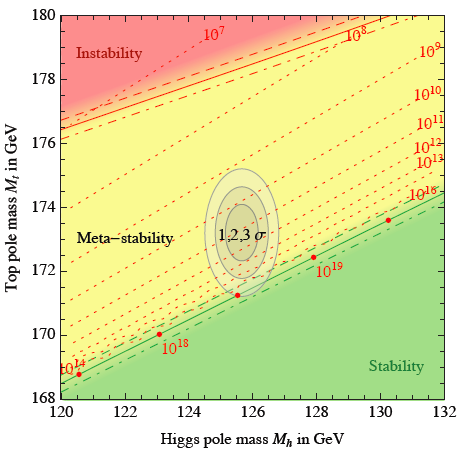}
\caption{ The regions of vacuum stability, metastability and instability in the $(m_H, m_t)$ plane, from~\protect\cite{Buttazzo}.}
\label{fig:Buttazzo}
\end{figure}

Calculations in the Standard Model indicate that the brim turns down at a Higgs scale $\Lambda$ given by~\cite{Buttazzo}
\begin{equation}
\log_{10} \left( \frac{\Lambda}{{\rm GeV}} \right) \; = \; 11.3 + 1.0 \left(\frac{m_H}{{\rm GeV}} - 125.66 \right)
- 1.2 \left( \frac{m_t}{{\rm GeV}} - 173.10 \right) + 0.4 \left(\frac{\alpha_s(M_Z) - 0.1184}{0.0007} \right) \, .
\label{Buttazzo}
\end{equation}
Inserting the values (\ref{mt}) and (\ref{mH}), we find the estimate
\begin{equation}
\Lambda \; = \; 10^{10.5 \pm 1.1}~{\rm GeV}
\label{Lambda}
\end{equation}
(beware that the errors are not symmetric and Gaussian). This calculation is most sensitive to $m_t$, as seen in Fig.~\ref{fig:Buttazzo}.
If we use the new D0 value of (\ref{newmt}) reported at this conference,
we find that $\log_{10} (\Lambda/{\rm GeV})$ decreases by 2.0, but if we use the new CMS value
reported at this conference (\ref{newmt}) we find that $\log_{10} (\Lambda/{\rm GeV})$ increases by 1.6.
Hence we would welcome a more accurate value of $m_t$.

%%%%%%%%%%%%%%%%%%%%%%%%%%%%%%%%%%%%%%%%%%%%%%%%%%%%%%%%%%%%%%%%%%%%%%%%%
%%
%%   use this format to include an .eps figure into your paper
%%
%%%%%%%%%%%%%%%%%%%%%%%%%%%%%%%%%%%%%%%%%%%%%%%%%%%%%%%%%%%%%%%%%%%%%%%%%%%

One might have chosen to ignore this instability on the grounds that the lifetime of the vacuum
would (probably) be much longer than the age of the Universe, but this is made more difficult by the
suggestion from the cosmic microwave background that the Universe once had a much
higher energy density during an inflationary epoch~\cite{AK2}. Fluctuations during this epoch could have
enhanced the rate for transition out of the region of the local electroweak minimum and towards
the anti-de-Sitter region~\cite{Oops}. The subsequent evolution would have been described by the
Fokker-Planck equation, and one would expect the anti-de-Sitter region to dominate, though
one could perhaps argue that we have been lucky enough to survive in a non-anti-de-Sitter
pocket. This might not be so implausible if there was a non-minimal amount of inflation, and the
problem could perhaps be avoided altogether with a judicious choice of higher-dimensional
terms in the effective potential~\cite{Sher}. However, such cosmological arguments underline that
electroweak vacuum instability is a potential problem (pun intended).

\section{Elementary or Composite?}

\begin{figure}[htb]
\centering
\includegraphics[height=2.3in]{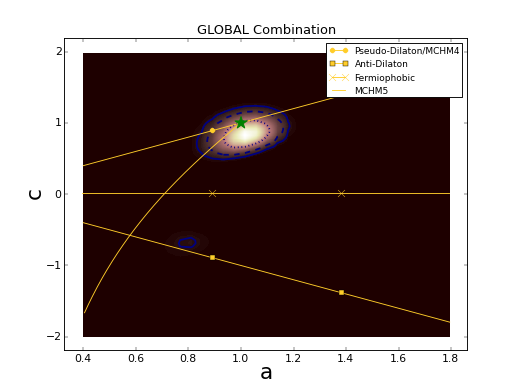}
\includegraphics[height=2.3in]{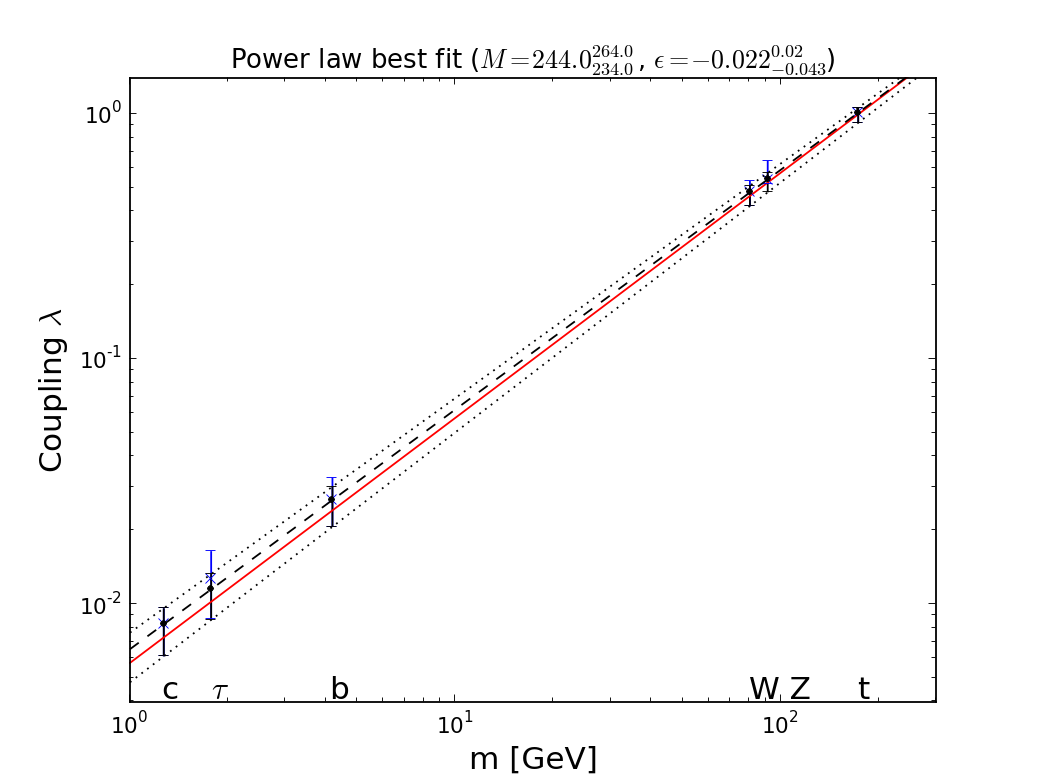}
\caption{ Left panel: a global fit to bosonic and fermionic $H$ couplings rescaled by factors $a$ and $c$, respectively,
indicating possible predictions of some composite models. Right panel: results from a two-parameter fit
(\protect\ref{Mepsilon}) to the $H$ couplings. Both plots are from~\protect\cite{EY3}.}
\label{fig:EY}
\end{figure}

The discussion in the previous Section assumed implicitly that the $H$ boson is
(effectively) elementary up to a large energy scale~\footnote{It was also assumed
that the spin of the $H$ boson is zero: by now simple alternatives are excluded with a very high
degree of confidence~\cite{WF,FM}.}. What evidence do we have that
this might be the case? It used to be thought that a composite Higgs boson would
normally have a mass comparable to the scale of compositeness, but this can be
reduced if it is a pseudo-Nambu-Goldstone boson whose mass is protected by some symmetry.
One way to probe such a possibility is to look for deviations from the Standard Model
predictions for the $H$ couplings. The left panel of Fig.~\ref{fig:EY} shows one particular example~\cite{EY3},
in which one looks for possible rescalings of the $H$ couplings to bosons by a factor $a$
and to fermions by a factor $a$: clearly there is no sign of a significant deviation from the
Standard Model case $a = c = 1$.

%%%%%%%%%%%%%%%%%%%%%%%%%%%%%%%%%%%%%%%%%%%%%%%%%%%%%%%%%%%%%%%%%%%%%%%%%
%%
%%   use this format to include an .eps figure into your paper
%%
%%%%%%%%%%%%%%%%%%%%%%%%%%%%%%%%%%%%%%%%%%%%%%%%%%%%%%%%%%%%%%%%%%%%%%%%%%%

Another test, this time of the dependence of $H$
couplings on particle masses, using a parameterization of the form
\begin{equation}
\lambda_f \; = \; \sqrt{2} \left(\frac{m_f}{M} \right), \; g_V \; = \; 2 \left( \frac{m_V^{2(1 + \epsilon)}}{M^{1 + 2 \epsilon}} \right)
\label{Mepsilon}
\end{equation}
for fermions and bosons, respectively. As seen in the right panel of Fig.~\ref{fig:EY},
the data are quite consistent with the Standard Model prediction $\epsilon = 0, M = 246$~GeV: we find
\begin{equation}
\epsilon \; = \; - 0.022^{+ 0.020}_{- 0.043}, \; M \; = 244^{+ 20}_{- 10}~{\rm GeV} \, .
\label{Mepsilonnumbers}
\end{equation}
For this reason, we wrote in one of our papers that the $H$ particle `walks and quacks like a Higgs boson'.
As we heard at this meeting, there are excellent prospects for testing the mass dependence of $H$
couplings, ranging from the muon to the top quark, with the energy and luminosity upgrades
of the LHC and possible future colliders, though the trilinear $H$ self-coupling may prove elusive~\cite{SD}.

\begin{figure}[htb]
\centering
\includegraphics[height=2.3in]{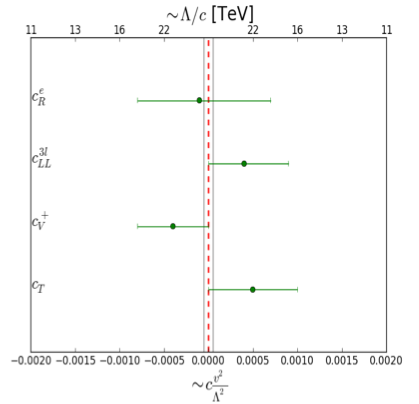}
\includegraphics[height=2.3in]{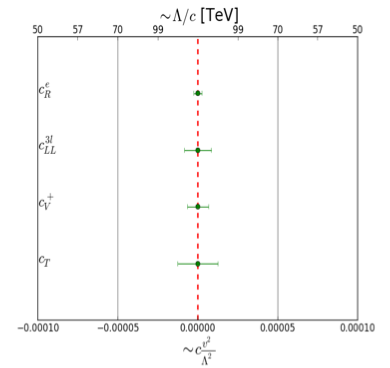}
\caption{ Left panel: current experimental constraints on the coefficients of dimension-6 operators~\protect\cite{dim6}.
 Right panel: possible future constraints from FCC-ee experiments~\protect\cite{FCC-ee}. The vertical `tram-lines' have the
 same locations in the two plots.}
\label{fig:ESY}
\end{figure}

A powerful and systematic way to probe for physics beyond the Standard Model is to
explore the constraints various experiments place on higher-dimensional interactions
that respect the Standard Model symmetries and might be generated in extensions of
the Standard Model~\cite{SD,MM,MN}. For example, composite models might generate such interactions
with coefficients depending inversely on the scale of compositeness. There are many such interactions already with
dimension 6, and many sources of constraints including precision electroweak measurements
as well as Higgs measurements. As an example, the latter can be used to constrain the combination
\begin{equation}
\Delta {\cal L} \; = \; \frac{c_T}{v^2} {\cal O}_T + \frac{c_V^+}{m_W^2} \left({\cal O}_W + {\cal O}_B \right) 
+ \frac{c_{LL}^{(3)\ell}}{v^2} {\cal O}_{LL}^{(3)\ell} \, ,
\label{dim6}
\end{equation}
where the operators ${\cal O}_{T,W,B}$ and ${\cal O}_{LL}^{(3)\ell}$ are defined in~\cite{dim6}.
The left panel of Fig.~\ref{fig:ESY} displays the current constraints on the coefficients $c_{T, W, B}$
and $c_{LL}^{(3)\ell}$ of these operators, and the right panel displays the prospective constraints
that might be placed by measurements at a possible large future circular $e^+ e^-$ collider (FCC-ee~\cite{FCC-ee}).
(Note the large difference in the horizontal scale!)

%%%%%%%%%%%%%%%%%%%%%%%%%%%%%%%%%%%%%%%%%%%%%%%%%%%%%%%%%%%%%%%%%%%%%%%%%
%%
%%   use this format to include an .eps figure into your paper
%%
%%%%%%%%%%%%%%%%%%%%%%%%%%%%%%%%%%%%%%%%%%%%%%%%%%%%%%%%%%%%%%%%%%%%%%%%%%%

\section{What Else is There? Supersymmetry!}

I consider supersymmetry to be the best-motivated possible accessible extension of the Standard Model~\cite{ML}.
In addition to all the familiar motivations for supersymmetry at the TeV scale, such as making the
hierarchy of mass scales more natural, providing an origin for the dark matter, facilitating grand
unification and playing an essential role in string theory, Run~1 of the LHC has provided three
additional motivations for supersymmetry. 1) It can stabilize the electroweak vacuum. 2) It predicted
the existence of a Higgs boson with a mass less than about 130~GeV, as measured~\footnote{Most
probably the LHC has discovered the lightest supersymmetric Higgs boson, but it is
also possible that the LHC may have discovered the second one, and that there is a lighter
supersymmetric Higgs boson still waiting to be discovered~\cite{lighter}.}. 3) It predicted
that the $H$ couplings would resemble those of the Standard Model Higgs boson, also as measured.

\begin{figure}[htb]
\centering
\includegraphics[height=2.3in]{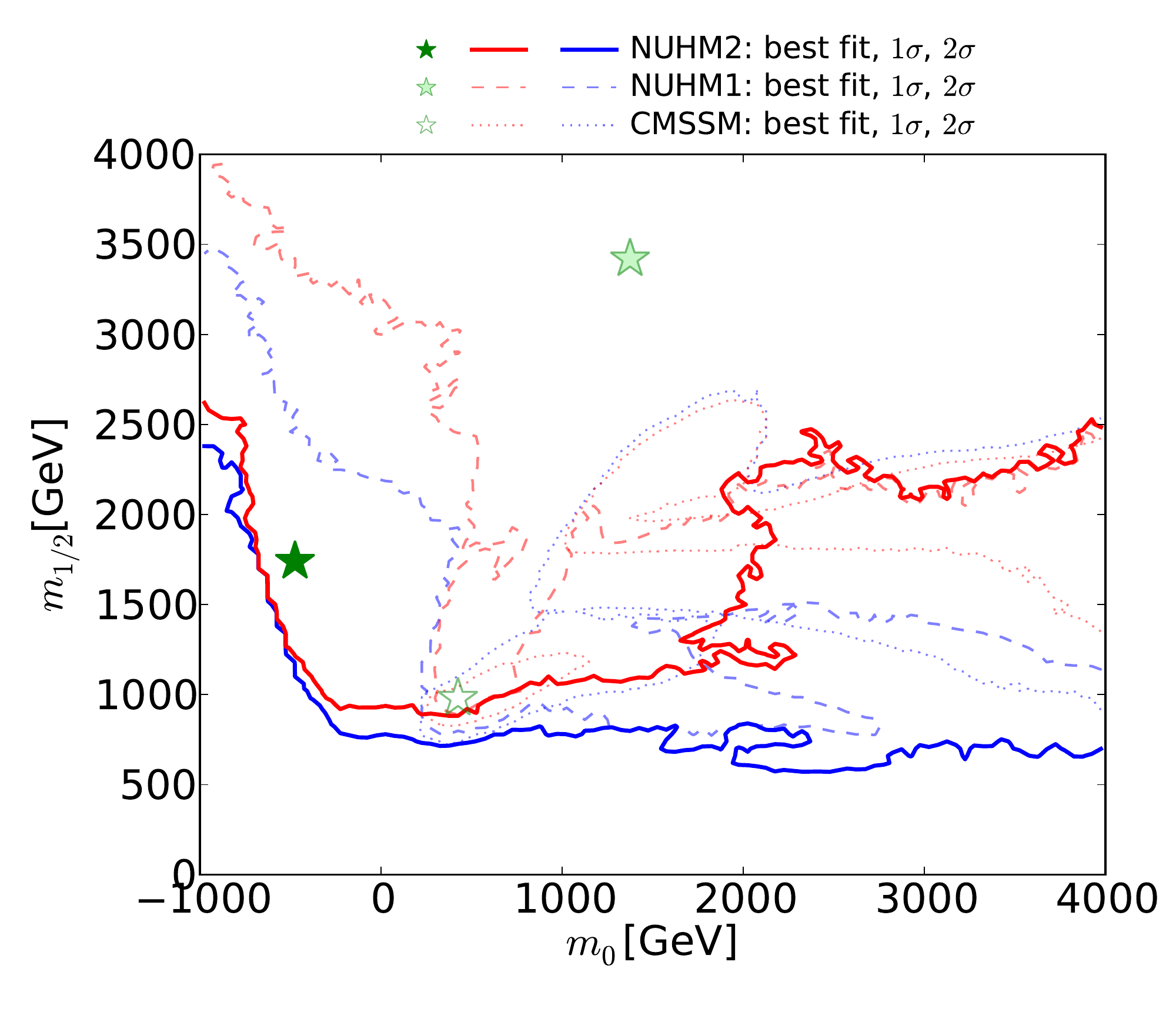}
\includegraphics[height=2.3in]{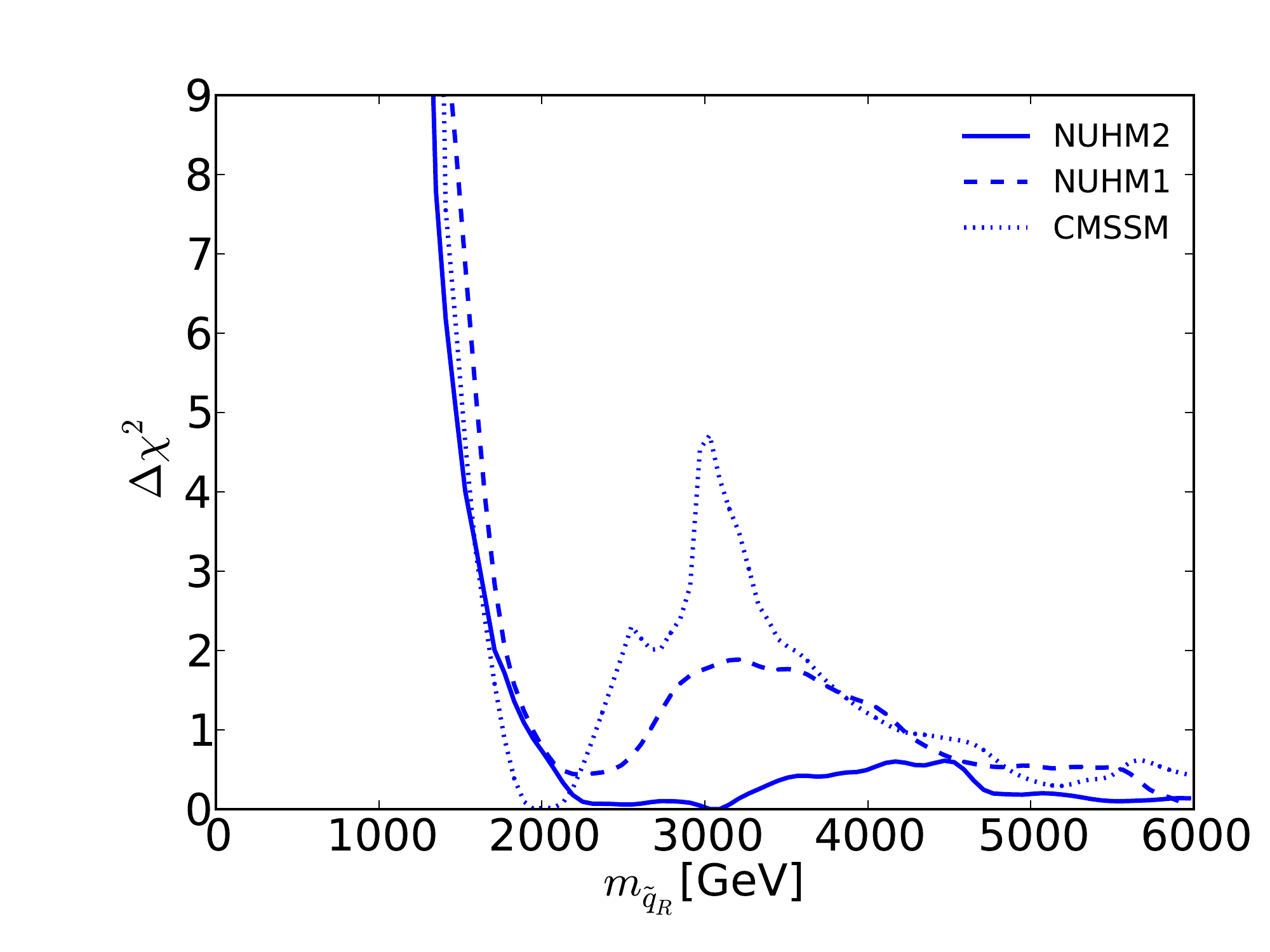}
\caption{ Left panel: the 68\% CL (red) and 95\% CL contours (blue) in the $(m_0, m_{1/2})$ panes
 for the CMSSM (dotted lines), NUHM1 (dashed lines) and NUHM2 (solid lines).
 Right panel: the one-dimensional $\chi^2$ likelihood function for the right-handed squark mass
 in these models. Both plots are from~\protect\cite{MC10}.}
\label{fig:MC10}
\end{figure}

However, despite all these motivations, direct searches for supersymmetry at the LHC have so
fare revealed nothing~\cite{MS,AM}, as is also the case with searches for the scattering of dark matter particles,
indirect searches in flavour physics, etc.. Combining all these constraints and requiring that the
supersymmetric relic density lie within the range required by cosmology for cold dark matter
imposes important bounds on the parameters of supersymmetric models, as seen in the left panel of Fig.~\ref{fig:MC10}
for the minimal supersymmetric extension of the Standard Model with the soft supersymmetry-breaking
parameters $(m_0, m_{1/2})$ constrained to be universal at the input GUT scale (the CMSSM), and in variants with
different soft supersymmetry-breaking contributions to the Higgs multiplet masses (NUHM1,2)~\cite{MC10}.
The right-hand panel of Fig.~\ref{fig:MC10} displays the
one-dimensional likelihood functions for the right-handed squark mass in the CMSSM and NUHM1,2.
The lower-mass regions allowed in these models could be explored by searches for
squarks and gluinos at the LHC. 

%%%%%%%%%%%%%%%%%%%%%%%%%%%%%%%%%%%%%%%%%%%%%%%%%%%%%%%%%%%%%%%%%%%%%%%%%
%%
%%   use this format to include an .eps figure into your paper
%%
%%%%%%%%%%%%%%%%%%%%%%%%%%%%%%%%%%%%%%%%%%%%%%%%%%%%%%%%%%%%%%%%%%%%%%%%%%%

\section{If you knows of a Better 'Ole, go to it!}

This was the caption of a famous cartoon from World War 1, and has some relevance
to the situation in particle physics today.
The LHC has presented us with a paradox: is there really nothing else in the electroweak sector
besides a single light Higgs boson? On the one hand, if there is something else
relatively light, why has it not shown up at the LHC, and why is there no indirect evidence for it? On the other hand, if there
really is nothing else light, is a light Higgs boson unnatural?

Some people take the point of view that the hierarchy problem is a red herring: it is a mathematical artefact of
one particular renormalization scheme that has no physical content. Others say that we just have to accept the
fine-tuning of the electroweak/Planck mass hierarchy: maybe there is an anthropic explanation, possibly
provided by the landscape of string theory. Others seek to solve, or at least rewrite the hierarchy problem
by postulating extra dimensions that may be warped. Yet others postulate new physics at the TeV scale
in the form of compositeness or supersymmetry~\cite{NC}.

Personally, I do not know of a better 'ole than supersymmetry.
In my view, the hierarchy problem is just as pressing than ever, and supersymmetry
is increasingly the best solution we have. As discussed above, Run~1 of the LHC has certainly not provided
any additional motivations for composite models or extra dimensions.
However, as discussed in the previous section, LHC Run 1
has provided new motivations for supersymmetry, but Nature is not described by simple
models unless the scale of supersymmetry breaking is a TeV or more. However, one
should regard supersymmetry as a framework, and not get hung up on specific models,
of which there are a large number, and many novel possibilities for supersymmetry remain.
Maybe R-parity is violated~\cite{NN}? Or maybe the spectrum is compressed?
Or maybe it is hidden or stealthy in some other way? Or maybe it is (semi-)split~\cite{NC}?
Run~2 of the LHC and beyond will certainly tighten the screws on possible scenarios
for physics beyond the Standard Model.

\section{The Future of Hadron Colliders}

Plans for future runs of the LHC include Run~2 with 30 to 100/fb at 13 or 14 TeV,
followed by Run~3, which will aim at 300/fb at 14~TeV~\cite{FB}. Plans for upgrades
of the LHC experiments are well underway~\cite{AR,AS,CG2,MM2}. Now included in CERN's
planning are preparations for a project to increase the
LHC luminosity further (HL-LHC), which aims at 3000/fb at 14~TeV~\cite{FB},
which will offer plenty of prospects for studying the Higgs boson~\cite{SS} and searching for new physics.
Beyond this, one possibility is to put high-field magnets in the LHC tunnel
(HE-LHC), which would aim at 3000/fb of collisions at 33 TeV in the centre of mass. More ambitious proposals
would involve digging new circular tunnels with circumferences between 50 and 70 km,as suggested
in China~\cite{CEPC}, or between 80 and 100~km,
as in the FCC-hh suggestion for CERN that would aim at 3000/fb of luminosity at 100 TeV~\cite{FB,FG}.

Detailed studies of such machines are starting now, and the physics agenda, which could include
high-luminosity $e^+ e^-$ collisions in the same tunnel~\cite{FCC-ee}, has yet
to be developed. However, it is already clear that such a machine would offer unique
prospects for Higgs physics, with over an order of magnitude more Higgs bosons
than HL-LHC in many channels, almost two orders of magnitude more for $t \bar{t} H$ production.
Such a collider would also extend greatly the physics reach for new particles, being able,
e.g., to discover gluinos weighing over 10~TeV~\cite{SARE}. Would such a machine be guaranteed to
find supersymmetry (if it exists), or to discover dark matter (if it was once in thermal
equilibrium in the early Universe)?

In the CMSSM and related models with R-parity, the dark matter density constraint typically allows very heavy
masses only along strips where the LSP is nearly degenerate with one or more other
sparticles, such as the lighter stau slepton or stop squark. The LHC will be able to
explore thoroughly the stau coannihilation strip in the CMSSM, but only part of the stop
coannihilation strip, which may extend to an LSP mass $\sim 6500$~GeV~\cite{EOZ}. Simple
scaling arguments indicate that in some cases
the stop coannihilation strip could be completely explored by the FCC-hh via standard
searches jets and missing transverse energy, or monojets, but more work on this question
is required.

\section{Patience!}

At the time of the Higgs discovery, {\it The Economist}, a well-known ``dismal science" journal, 
published a graphic illustrating the time-lags between the theoretical proposal and experimental
discovery of many Standard Model particles. In some cases, such as the muon, there was no time-lag,
because theorists had not thought of the particle beforehand:
as the famous Columbia University physicist I.~I.~Rabi famously remarked `` Who ordered that?".
The longest time-lag was that for the Higgs boson: 48 years between proposal and discovery.
Lovers of supersymmetry should be patient: four-dimensional supersymmetric field theories
were first published in 1974: only 40 years so far. If Run~2 of the LHC discovers
supersymmetry, the time-lag will still have been less than for the Higgs boson!

%%  if necessary
\Acknowledgements
Work supported by the European Research Council 
via the Advanced Investigator Grant 267352 and by STFC (UK) via the research grant ST/J002798/1.

\end{document}